\documentclass[aps,prl,preprint,unsortedaddress]{revtex4-1}
\usepackage[dvips]{graphicx}

\begin{document}

\title{Gibbs-Jaynes Entropy Versus Relative Entropy}

\author{M. Mel\'endez}
\affiliation{Dpto. F\'isica Fundamental, UNED\\
             Senda del Rey, 9\\
             28040 Madrid}
\email{mmelendez@fisfun.uned.es}

\author{P. Espa\~nol}
\affiliation{Dpto. F\'isica Fundamental, UNED\\
             Senda del Rey, 9\\
             28040 Madrid}

\date{\today}

\begin{abstract}
The maximum entropy formalism developed by Jaynes determines the relevant
ensemble in nonequilibrium statistical mechanics by maximising the entropy
functional subject to the constraints imposed by the available information. We
present an alternative derivation of the relevant ensemble based on the
Kullback-Leibler divergence from equilibrium. If the equilibrium ensemble is
already known, then calculation of the relevant ensemble is considerably
simplified. The constraints must be chosen with care in order to avoid
contradictions between the two alternative derivations. The relative entropy
functional measures how much a distribution departs from equilibrium. Therefore,
it provides a distinct approach to the calculation of statistical ensembles
that might be applicable to situations in which the formalism presented by
Jaynes performs poorly (such as non-ergodic dynamical systems).

[The final publication is available at Springer via
http://dx.doi.org/10.1007/s10955-014-0954-6]
\keywords{Gibbs-Jaynes entropy \and Kullback-Leibler divergence
          \and Relative entropy \and Maximum entropy formalism
          \and Nonequilibrium statistical mechanics}
\end{abstract}

\maketitle

\section{Introduction}
\label{intro}

Edwin T. Jaynes wrote a beautiful article in 1957 \cite{Jaynes-1957} advocating
a reinterpretation of statistical mechanics in light of Shannon's mathematical
theory of communication \cite{Shannon}. Instead of working with ensembles of
systems, Jaynes posed the problem in the following terms: suppose we know the
expected values of a set of functions $F_1,\ \ldots,\ F_k$ of the microscopic
state $z$ of a system, what is the best estimate for the average value of some
other function $G(z)$? Without access to $z$, which is never available in the
lab, the best that can be done is to pick a probability distribution $\rho (z)$
over the states and then calculate the expected value of $G$ as
$\mathrm{Tr}\left[\rho G\right]$. But then we encounter the problem of which
distribution $\rho$ to choose, because the average values
$\left<F_1\right>,\ \ldots,\ \left<F_k\right>$ do not provide enough information
to determine $\rho$ uniquely. We need an additional criterion. Therefore,
argued Jaynes, one  should use the distribution that maximises the Shannon
entropy functional
\begin{equation}
\mathrm{H}[\rho] \equiv -\sum_ i\rho(z_i) \ln(\rho(z_i)),
\end{equation}
subject to the constraints imposed by the information available. Any other
distribution would imply an unjustified bias in the probabilities.

Shannon's $\mathrm{H}$ functional applies only to discrete distributions, but
the Gibbs-Jaynes entropy functional $S$ is analogous to $\mathrm{H}$ for
continuous sets of states, as in the case of points in phase space
\cite{Jaynes-1963}.
\begin{equation}
\label{S}
S[\rho] \equiv -k_B \int_{\Gamma}\rho(z)\ln\left(\frac{\rho(z)}{m(z)}\right)\ dz
  = -k_B \mathrm{Tr}\left[\rho \ln\left(\frac{\rho}{m}\right)\right].
\end{equation}
In classical Hamiltonian dynamics, the measure $m(z)$ turns out to be
$h^{-3N}$ ($N$ is the number of particles and $h$ is Planck's constant).
$\Gamma$ stands for the whole phase space and $k_B$ for Boltzmann's
constant.

Jaynes's maximum entropy formalism allows us not only to derive equilibrium
statistical mechanics from the point of view of statistical inference, but also
to select probability distributions in more general situations, when the
expected values of several arbitrary phase functions have been established, even
if they are not dynamical invariants. Working out such distributions constitutes
the key step for projection operator techniques in the theory of transport
processes \cite{Kawasaki,Grabert,Zubarev}.

Recent developments in nonequilibrium statistical mechanics
\cite{Gaveau,Qian,Kawai,Shell,Vaikuntanathan,Horowitz,Roldan,Crooks-Sivak,Crooks,Sivak-Crooks}
have borrowed another tool from information theory known as the Kullback-Leibler
divergence \cite{Kullback-Leibler},
\begin{equation}
D(\rho\|\rho') \equiv
   \int_{\Gamma} \rho(z) \ln\left(\frac{\rho(z)}{\rho'(z)}\right)\ dz,
\end{equation}
which measures how ``different'' $\rho$ is from $\rho'$. A well-known
result in information theory states that $D(\rho\|\rho') \geq 0$, with
equality holding only when $\rho(z) = \rho'(z)$ almost everywhere.

When an equilibrium ensemble is used as the reference distribution,
$\rho' = \rho^{eq}$, there is a simple connection between the Gibbs-Jaynes
entropy (\ref{S}) and the \textit{relative entropy}, defined here as
\begin{equation}\label{Def-Relative_entropy}
\Delta S[\rho] \equiv -k_B\ D(\rho\|\rho^{eq}).
\end{equation}
The link can easily be established the moment we realise that the equilibrium
ensemble is a stationary solution of Liouville's equation,
\begin{equation}
\frac{\partial \rho^{eq}}{\partial t} = 0,
\end{equation}
and that $\rho^{eq}$ must therefore be a function of the dynamical invariants
only. Let $I(z)$ stand for the invariants for the microstate $z$, such as the
energy, the linear momentum, and so forth. We then know that $\rho^{eq}(z)$ is
some function $\Phi(I(z))$. The integral of $\rho^{eq}$ over all the states that
satisfy $I(z) = i$ should equal the probability $P(i)$ of finding the system in
a state compatible with these values of the invariants.
\begin{equation}
\mathrm{Tr}\left[\rho^{eq}\delta(I-i)\right] =
  \Phi(i)\mathrm{Tr}\left[\delta(I-i)\right] = P(i),
\end{equation}
where $\delta$ is the Dirac delta function. Solving for $\Phi$, we end up with
\begin{equation}
\label{Phi}
\Phi(i) = \frac{P(i)}{\mathrm{Tr}\left[\delta(I-i)\right]},
\end{equation}
which means that
\begin{equation}\label{rho_eq}
\rho^{eq}(z) = \frac{P(I(z))}{h^{3N}\Omega(I(z))}.
\end{equation}
The function $\Omega(i)$ in the denominator may be thought of as the ``number of
microstates'' that satisfy $I(z) = i$,
\begin{equation}
\Omega(i) = \int_{\Gamma} h^{-3N}\delta(I(z)-i)\ dz.
\end{equation}
When the expression for the equilibrium distribution (\ref{rho_eq}) is
substituted into the definition of the relative entropy
(\ref{Def-Relative_entropy}), we find
\begin{eqnarray}
\label{Delta_S_theorem}
\Delta S[\rho] & = &
 -k_B \int_{\Gamma} \rho(z) \ln\left(\frac{h^{3N}\rho(z)}
                                          {\frac{P(I(z))}
                                                      {\Omega(I(z))}}\right)\ dz
\\ & = &
  -k_B \int_{\Gamma} \rho(z) \ln\left(h^{3N}\rho(z)\right)\ dz
  +k_B \int_{\Gamma} \rho(z) \ln\left(\frac{P(I(z))}{\Omega(I(z))}\right)\ dz
\nonumber \\
   & = &
  S[\rho] + k_B \int \mathrm{Tr}\left[\rho \delta(I-i)\right]
                     \ln\left(\frac{P(i)}{\Omega(i)}\right)\ di, \nonumber
\end{eqnarray}
(the latter integral is extended over all the values of $i$). Hence maximising
the relative entropy functional $\Delta S$ is equivalent to maximising the
Gibbs-Jaynes functional $S$ \textit{as long as this last integral is constant
for the distributions allowed by the constraints}. A reasonable way to meet this
condition requires
\begin{equation}\label{probability_assumption}
\mathrm{Tr}\left[\rho \delta(I-i)\right] = P(i).
\end{equation}
In other words, the agreement between both maximisation strategies is based on
the assumption that the unknown distribution generates the same probability
distribution over the dynamical invariants as the equilibrium ensemble. If
equation (\ref{probability_assumption}) is conceded then, given the probability
distribution at equilibrium, we can follow two different paths to calculate the
least biased distribution compatible with our information about the system.

An anonymous reviewer pointed out that condition (\ref{probability_assumption})
could be relaxed when $\rho^{eq}$ designates the canonical ensemble
(\ref{canonical_ensemble}), for then we can simply assume that the expected
energy calculated with $\rho$ leads to the same result as when it is calculated
with $\rho^{eq}$ and then prove that the final integral in equation
(\ref{Delta_S_theorem}) becomes independent of $\rho$.
\begin{equation}
-k_B \int_{\Gamma} \rho(z) \ln\left(h^{3N}\rho^{eq}(z)\right)\ dz.
 =
-k_B \int_{\Gamma} \rho^{eq}(z) \ln\left(h^{3N}\rho^{eq}(z)\right)\ dz
 =
S\left[\rho^{eq}\right]
\end{equation}
Hence, in the canonical case condition (\ref{probability_assumption}) may be
relaxed to
\begin{equation}
\mathrm{Tr}[\rho I] = \mathrm{Tr}[\rho^{eq} I]
\end{equation}
Similarly, for a microcanonical $\rho^{eq}$, it is enough to require that
$\rho$ vanishes whenever $\rho^{eq}$ does.

\section{Generalised canonical probability distribution}
\label{g.c.d.}

Let us illustrate the two alternative maximisation routes by working out the
classic example of the probability distribution for a system in contact with a
heat bath at temperature $T$. We shall start with the Gibbs-Jaynes functional
and introduce two habitual assumptions. First, we will imagine that the system
and the reservoir have been isolated from the rest of the universe, and second,
we will disregard the interaction energy between them, which we assume is very
small compared with their internal energies, so that the Hamiltonian, which
remains constant, may be expressed as a sum of two terms,
\begin{equation}\label{Hamiltonian}
H(z) = H_S(z_S) + H_R(z_R),
\end{equation}
the former corresponding to our system, and the latter to the reservoir.

A typical setup might also include the measured values of several
macroscopic variables, such as concentrations or hydrodynamic velocity fields
\cite{Grabert}. We denote these values $f_1,\ \ldots,\ f_k$ and equate them to
the averages for the corresponding functions
$\left<F_1\right>,\ \ldots,\ \left<F_k\right>$, of the microstate $z_S$
\begin{equation}
\label{lambda_constraints}
f_i = \mathrm{Tr}\left[\rho F_i\right],
\end{equation}
where $\rho$ is the unknown distribution. Given the one-to-one correspondence
between the equilibrium temperature $T$ and the total internal energy, we
include an additional constraint for the expected value of $H(z)$,
\begin{equation}
E = \mathrm{Tr}\left[\rho H\right],
\end{equation}
although we do not yet know the actual number represented by $E$. Finally, we
must ensure that $\rho$ is properly normalised,
\begin{equation}
\label{normalisation}
\mathrm{Tr}\left[\rho\right] = 1.
\end{equation}

\subsection{Gibbs-Jaynes entropy functional}
\label{S_route}

We now wish to find the distribution that maximises the entropy functional
subject to all the constraints (\ref{lambda_constraints})-(\ref{normalisation}).
Following the standard method of Lagrange multipliers, we add constraint terms
to the Gibbs-Jaynes functional (\ref{S}) to obtain
\begin{eqnarray}
C[\rho] & = &
    -k_B\mathrm{Tr}\left[\rho\ \ln\left(h^{3N}\rho\right)\right]
    -k_B\sum_{i=1}^k\lambda_i\left(\mathrm{Tr}\left[\rho F_i\right]-f_i\right)
\\ &  &
    -k_B\beta\left(\mathrm{Tr}\left[\rho H\right] - E\right)
    -k_B\mu\left(\mathrm{Tr}\left[\rho\right] - 1 \right). \nonumber
\end{eqnarray}
The $\lambda_i$, $\beta$ and $\mu$ are all Lagrange multipliers. The
functional derivative of $C$ with respect to $\rho$ should vanish for the
least biased distribution, frequently called the \textit{relevant}
distribution, $\bar{\rho}$,
\begin{equation}
\left.\frac{\delta C}{\delta\rho}\right|_{\rho=\bar{\rho}} = 0,
\end{equation}
and this equation leads us to
\begin{equation}
\label{rel_rho}
\bar{\rho} = h^{-3N}e^{-\sum_{i=1}^k \lambda_i F_i -\beta H -\mu - 1}.
\end{equation}
Substitution of $\bar{\rho}$ into the constraints should allow us, in principle,
to calculate the Lagrange multipliers. Equation (\ref{normalisation}), for
example, determines the value of $\mu$. When combined with (\ref{rel_rho}),
we find that
\begin{equation}
e^{-\mu - 1} = \frac{1}{\mathrm{Tr}\left[h^{-3N}e^{-\sum_{i=1}^k \lambda_i F_i
                                                 -\beta H}\right]}
            = \frac{1}{Z},
\end{equation}
so our probability distribution becomes
\begin{equation}\label{total_rho_bar}
\bar{\rho}(z) = \frac{1}{h^{3N}Z} e^{-\sum_{i=1}^k\lambda_i F_i(z_S)
                                     -\beta H(z)}.
\end{equation}
Equation (\ref{Hamiltonian}) implies that the partition function $Z$ factors
into the product of an integral over $z_S$ and another over $z_R$. We will
refer to these factors as $Z_S$ and $Z_R$.
\begin{eqnarray}\label{Z1Z2}
Z & = & \left(
        \int_{\Gamma_S} h^{-3N_S} e^{-\sum_{i=1}^k\lambda_i F_i(z_S)
                                    -\beta H_S(z_S)}\ dz_S
        \right)
        \left(
        \int_{\Gamma_R} h^{-3N_R} e^{-\beta H_R(z_R)}\ dz_R
        \right) \\
  & = & Z_S Z_R, \nonumber
\end{eqnarray}
where $N_S$ and $N_R$ are the number of particles in the system and reservoir,
respectively.

We define the entropy $S(E,\ f_1, \ldots,\ f_k)$ as the maximum value of the
Gibbs-Jaynes entropy functional. Inserting the expression for $\bar{\rho}$ into
(\ref{S}) reveals the following link between the partition function and the
entropy:
\begin{equation}
\label{S_at_rho_bar}
S(E,\ f_1,\ \ldots,\ f_k) \equiv S\left[\bar{\rho}\right]
        =   k_B\ln(Z) + k_B\sum_{i=1}^k \lambda_i f_i
                      + k_B\beta E.
\end{equation}
Factoring $Z$ according to (\ref{Z1Z2}), the entropy separates neatly into two
terms,
\begin{equation}\label{S_is_extensive}
S(E,\ f_1,\ \ldots,\ f_k) =
    k_B\left(\ln(Z_S) + \sum_{i=1}^k \lambda_i f_i + \beta E_S \right)
 +  k_B\left(\ln(Z_R) + \beta E_R \right)
\end{equation}
where $E_S$ and $E_R$ stand for the expected values of $H_S$ and $H_R$. On the
right of equation (\ref{S_is_extensive}) we find the entropies of the system and
reservoir considered separately. In other words, suppose we had isolated the
system from the heat bath and had then calculated the entropies for both
independently by maximising the Gibbs-Jaynes functional, with the same
constraints on the average values of $F_1,\ \ldots,\ F_k$ and the normalisation
of $\rho$, but changing the constraints on the expected energies to
$\left<H_S\right>_S = E_S$ for the system and $\left<H_R\right>_R = E_R$ for the
bath, where $E_S$ and $E_R$ are the same values as in equation
(\ref{S_is_extensive}). Then the expressions for the entropies, $S_S$ for the
system and $S_R$ for the reservoir, would read
\begin{eqnarray}\label{SS_and_SR}
S_S(f_1,\ \ldots,\ f_k,\ E_S) & = & k_B\left(\ln(Z_S)
                                             + \sum_{i=1}^k \lambda_i f_i
                                             + \beta_S E_S \right), \\
S_R(E_R) & = &  k_B\left(\ln(Z_R) + \beta_R E_R \right). \nonumber
\end{eqnarray}
We will show below that the sum $S_S + S_R$ equals $S$ in (\ref{S_is_extensive})
because
\begin{equation}\label{thermal_equilibrium}
\beta_S = \beta_R = \beta.
\end{equation}
The set of equations (\ref{S_is_extensive}), (\ref{SS_and_SR}) and
(\ref{thermal_equilibrium}) illustrates the well-known fact that entropy is an
extensive quantity if the interaction between subsystems is small enough to be
disregarded. The equality of the three Lagrange multipliers in
(\ref{thermal_equilibrium}) can be verified by comparing the average values
calculated using (\ref{total_rho_bar}) to the averages for the system and
reservoir considered independently, noting that
\begin{eqnarray}
\left<H_S\right>_S & = & \left<H_S\right> = E_S, \\
\left<H_R\right>_R & = & \left<H_R\right> = E_R. \nonumber
\end{eqnarray}

The Lagrange multiplier $\beta_R$ is related to the temperature of the heat
bath according to
\begin{equation}\label{T}
  \frac{1}{T} = \frac{\partial S_R}{\partial E_R} = k_B \beta_R.
\end{equation}
Therefore, equation (\ref{thermal_equilibrium}) shows that the temperatures of
the system and the bath must equal the same value $T$ for the relevant
distribution, and
\begin{equation}
  \beta = \frac{1}{k_B T}.
\end{equation}
Equation (\ref{total_rho_bar}) then becomes
\begin{equation}
\bar{\rho}(z) = \frac{1}{h^{3N} Z} e^{-\sum_{i=1}^k \lambda_iF_i(z_S)
                                      -\frac{1}{k_BT}H_S(z_S)
                                      -\frac{1}{k_BT}H_R(z_R)}.
\end{equation}
To calculate the probability distribution for $z_S$, we can now integrate
over the degrees of freedom of the reservoir, that is,
\begin{equation}
\label{integral_over_z_R}
\bar{\rho}_S(z_S) = \int \bar{\rho}(z)\ dz_R.
\end{equation}
The answer to our problem, known as the generalised canonical probability
distribution \cite{Grabert}, is the least biased probability distribution for a
system at constant temperature $T$ that also satisfies the constraints
(\ref{lambda_constraints}). Carrying out the integral
(\ref{integral_over_z_R}), we obtain
\begin{equation}\label{gcd}
\bar{\rho}_S(z_S) = \frac{e^{-\sum_{i=1}^k\lambda_iF_i(z_S)
                             -\frac{1}{k_BT}H_S(z_S)}}
                     {\mathrm{Tr}\left[e^{-\sum_{i=1}^k\lambda_iF_i(z_S)
                                          -\frac{1}{k_BT}H_S(z_S)}\right]},
\end{equation}
where the trace should now be interpreted as an integration over $z_S$.

\subsection{Relative entropy functional}

Now consider the derivation of the generalised canonical distribution
(\ref{gcd}) from the relative entropy (\ref{Def-Relative_entropy}).
In that case, we do not need to pay attention to the reservoir, so we maximise
the functional $\Delta C$, which includes the relative entropy and constraints
(\ref{lambda_constraints}) and (\ref{normalisation}).
\begin{equation}
\Delta C = 
    -k_B\mathrm{Tr}\left[\rho\ \ln\left(\frac{\rho}{\rho^{eq}}\right)\right]
    -k_B\sum_{i=1}^k\lambda_i\left(\mathrm{Tr}\left[\rho F_i\right]-f_i\right)
    -k_B\mu\left(\mathrm{Tr}\left[\rho\right] - 1 \right).
\end{equation}
As before, we calculate the functional derivative of $\Delta C$ with respect
to $\rho$ and find that
\begin{equation}\label{rel_rho_2}
\bar{\rho} = \rho^{eq}e^{-\sum_{i=1}^k\lambda_i F_i -\mu - 1}.
\end{equation}
The equilibrium distribution $\rho^{eq}$ for a system at constant temperature
is the well-known canonical ensemble,
\begin{equation}\label{canonical_ensemble}
\rho^{eq}(z_S) = \frac{e^{-\frac{1}{k_BT}H_S(z_S)}}
                       {\mathrm{Tr}\left[e^{-\frac{1}{k_BT}H_S(z_S)}\right]}.
\end{equation}
By ensuring that $\bar{\rho}$ is normalised, we determine $\mu$
\begin{equation}\label{normalisation_factor}
e^{-\mu - 1} = \frac{\mathrm{Tr}\left[e^{-\frac{1}{k_BT}H_S(z_S)}\right]}
                    {\mathrm{Tr}\left[e^{-\sum_{i=1}^k \lambda_i F_i(z_S)
                                         -\frac{1}{k_BT}H_S(z_S)}\right]}.
\end{equation}
And substituting (\ref{canonical_ensemble}) and (\ref{normalisation_factor})
into (\ref{rel_rho_2}), we recover the generalised canonical probability
distribution (\ref{gcd}).

Jaynes's maximum entropy formalism guided us to the desired solution, but the
path we had to follow was not as direct as the relative entropy route.
Furthermore, in the former derivation, we found ourselves describing the effect
of the reservoir in terms of the total energy $E = E_S + E_R$. But knowledge of
the energy in the reservoir, $E_R$, clearly has no bearing on our problem
because heat baths are characterised by their temperature, not their internal
energy, and it is a good thing that $E_R$ eventually drops out of the equations.
Therefore, if we already know the equilibrium ensemble, perhaps it is easier to
derive the relevant distribution from the relative entropy functional.
Nevertheless, the functional form of equation (\ref{gcd}) could have been
inferred from the Gibbs-Jaynes entropy (\ref{S}) by a simpler procedure that
does not contemplate the reservoir. The idea is to use the constraints for the
average values $f_i$ (\ref{lambda_constraints}), normalisation
(\ref{normalisation}) and an extra constraint for the expected value of the
unknown energy $E_S$. The maximum entropy formalism then leads to an expression
analogous to (\ref{gcd}), but with an unknown coefficient before $H_S(z_S)$.
All the extra work with the reservoir in the subsection on the Gibbs-Jaynes
derivation was carried out to establish that the temperature in equation
(\ref{gcd}) was equal to the temperature of the reservoir $T$ (note that we have
not assumed thermal equilibrium between the reservoir and system of interest).
When the same result was derived from the relative entropy functional
(\ref{Def-Relative_entropy}) we did not have to do any of this extra work
because the relevant information was already captured in the equilibrium
distribution.

\section{An apparent paradox}
\label{paradox}

The preceding discussion might give the impression that the relevant
distribution may always be expressed as
\begin{equation}\label{relevant_eq}
\bar{\rho}(z) = \frac{\rho^{eq}(z)}{\mathcal{Z}} e^{\sum_i \lambda_i F_i(z)},
\end{equation}
where $\mathcal{Z}$ stands for the appropriate normalisation factor. But this
rule may lead to incorrect conclusions if applied carelessly. To see why, let
us examine a slightly more general problem.

Whenever we are dealing with macroscopic systems in experiments, the exact
number of atoms or molecules remains unknown. Let $z_N$ represent the
coordinates and momenta of a system of $N$ particles. The probability
distributions and functions will now depend on the dimensionality of phase
space, so we will write them with a subindex $N$ to emphasise this dependence.
The Gibbs-Jaynes entropy functional (\ref{S}) can be generalised to
\begin{equation}\label{Jaynes_entropy}
S=-k_B \sum_N \int_{\Gamma_N} \rho_N(z_N)\
                              \ln\left(h^{3N}\rho_N(z_N)\right)\ dz_N,
\end{equation}
and the constraints on the average values $\left<F_1\right>,\ \ldots,\ 
\left<F_k\right>$ now read
\begin{equation}\label{macro_f_constraints}
f_i = \sum_N \int_{\Gamma_N} \rho_N(z_N)F_{i,N}(z_N)\ dz_N,
\end{equation}
where $\Gamma_N$ represents the phase space for $N$ particles. Similarly, the
expression for the relative entropy turns into
\begin{equation}\label{relative_entropy}
\Delta S=-k_B \sum_N \int_{\Gamma_N}\rho_N(z_N)\ 
                                    \ln\left(\frac{\rho_N(z_N)}
                                               {\rho_N^{eq}(z_N)}\right)\ dz_N.
\end{equation}
The obvious generalisation of (\ref{relevant_eq}) must be
\begin{equation}\label{macrocanonical_rho_eq}
\bar{\rho}_N(z_N) = \frac{\rho_N^{eq}(z_N)}
                       {\mathcal{Z}}     e^{\sum_i \lambda_i F_{i,N}(z_N)}.
\end{equation}
Note that $\rho_N(z_N)$ and $\rho^{eq}_N(z_N)$ represent joint
probability densities for $N$ and $z_N$, and are therefore not normalised to
one, but rather
\begin{equation}
\int_{\Gamma_N} \rho_N(z_N)\ dz_N = \int_{\Gamma_N} \rho^{eq}_N(z_N)\ dz_N
                                  = P(N),
\end{equation}
where P(N) represents the probability of $N$ particles in the system.

Imagine an isolated system for which we know the probability distribution
$P(E,\ N)$ for the total energy $E$ and number of particles $N$. Both $E$ and
$N$ are dynamical invariants, and so is the probability $P(E,\ N)$. Hence we
should find the same probability for $E$ and $N$ at equilibrium.
\begin{eqnarray}\label{probability_constraint}
P(E,\ N) & = & \int_{\Gamma_N} \rho_N(z_N) \delta(H_N(z_N) - E)\ dz_N \\
         & = & \int_{\Gamma_N} \rho_N^{eq}(z_N) \delta(H_N(z_N) - E)\ dz_N.
               \nonumber
\end{eqnarray}
If we use Lagrange's method to maximise (\ref{Jaynes_entropy}) subject to
(\ref{macro_f_constraints}) and (\ref{probability_constraint}), we derive the
set of relevant distributions
\begin{equation}\label{Jaynes_solution}
\bar{\rho}_N(z_N) = P(H_N(z_N),\ N)
                    \frac{e^{\sum_i \lambda_i F_{i,N}(z_N)}}
                         {\int_{\Gamma_N} e^{\sum_i \lambda_i F_{i,N}(z_N')}
                               \delta(H_N(z_N') - H_N(z_N))\ dz_N'}.
\end{equation}
But in general these functions are formally different from our previous
expression for $\bar{\rho}_N$ (\ref{macrocanonical_rho_eq}).

The disagreement between the two methods dissolves when we carry out the
operations carefully. It might seem at first that there is no need to
include the constraint (\ref{probability_constraint}) when we maximise the
functional for the relative entropy, because all the relevant information
about $P(E,\ N)$ should already be included in the equilibrium distribution.
However, we do in fact have to specify that the distribution we are looking for
must lead to the same value as the equilibrium distribution when both are
integrated over a given constant energy manifold. When using the relative
entropy (\ref{relative_entropy}), the correct functional to maximise is thus
\begin{eqnarray}
\Delta C & = & \Delta S
-k_B\sum_i\lambda_i
          \left(\sum_N\int_{\Gamma_N}\rho_N(z_N)F_{i,N}(z_N)\ dz_N-f_i\right)
\\       &   &
-k_B\sum_N\int_0^\infty\mu(E, N)
\left(\int_{\Gamma_N}\rho_N(z_N)\delta(H_N(z_N)-E)\ dz_N-P(E,\ N)\right)\ dE.
\nonumber
\end{eqnarray}
The last term above includes a Lagrange multiplier $\mu$ for each pair of $E$
and $N$, as required by equation (\ref{probability_constraint}). Once again, we
equate the derivatives of $\Delta C$ to zero and solve for the relevant
distribution to find
\begin{equation}\label{relative_solution}
\bar{\rho}_N(z_N)=\rho_N^{eq}(z_N)
                  e^{-\mu(H_N(z_N),\ N) - 1 -\sum_i \lambda_i F_{i,N}(z_N)}.
\end{equation}
This distribution can be identified with (\ref{macrocanonical_rho_eq}) as long
as $\mathcal{Z}$ is interpreted as a function of $E$ and $N$. In other words, if
we define
\begin{equation}\label{Def-mathcal_Z}
\mathcal{Z}(E,\ N) = e^{\mu(E,\ N) + 1},
\end{equation}
then we can simply insert (\ref{relative_solution}) into
(\ref{probability_constraint}) and solve for $\mathcal{Z}$, to determine
\begin{equation}\label{mathcal_Z}
\mathcal{Z}(E,\ N) = \frac{\int_{\Gamma_N} \rho_N^{eq}(z_N)
                                           e^{-\sum_i\lambda_iF_{i,N}(z_N)}
                                           \delta(H_N(z_N) - E)\ dz_N}
                          {P(E,\ N)}.
\end{equation}
Recalling the expression for the equilibrium distribution (\ref{rho_eq}),
\begin{equation}\label{rho_eq_N}
\rho_N^{eq}(z_N) = \frac{P(H_N(z_N),\ N)}{h^{3N}\Omega_N(H_N(z_N))},
\end{equation}
equations (\ref{Def-mathcal_Z})-(\ref{rho_eq_N}) can be used to convert
(\ref{relative_solution}) into (\ref{Jaynes_solution}), so the two methods once
again lead to the same result, as they should.

\section{Preliminary results concerning applications}
\label{example}

In the literature, relative entropy has been applied mainly to the calculation
of nonequilibrium free energy differences \cite{Gaveau,Qian,Sivak-Crooks} and
dissipated work \cite{Kawai}. In that context, $\rho$ and $\rho^{eq}$ are both
distributions that can be realised physically, such as the equilibrium
ensembles of a given Hamiltonian. By contrast, in the present paper we have
used the relative entropy functional to determine \textit{relevant
distributions}, which need not be realised physically, because they represent the
least-biased distribution that is consistent with the information available. 

\begin{figure}
  \centering
    \includegraphics[height=0.4\textwidth]{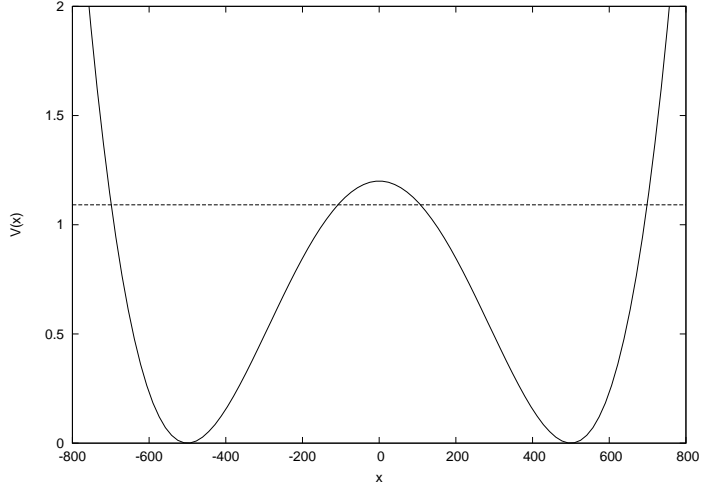}
    \caption{\label{doublewell} Double-well potential confining $N = 100$
      Lennard-Jones particles. The dashed line marks the average energy per
      particle $E/N$.}
\end{figure}

Within the theory of Mori-Zwanzig transport processes, relevant distributions
have become a crucial tool to derive generalised transport equations
\cite{Kawasaki,Grabert,Zubarev}. Consider the one-dimensional isolated
double-well potential drawn in figure \ref{doublewell}, which confines one
hundred particles that interact with each other through the Lennard-Jones
potential. Let the relevant variable $F(z)$ represent the number of particles
on the right,
\begin{equation}
\label{relevant_variable}
F(z) = \sum_{i = 1}^N \theta(q_i),
\end{equation}
where $\theta$ is the Heaviside step function and $q_i$ the position of
particle $i$. The Mori-Zwanzig theory of nonequilibrium transport allows us to
write exact transport equations for the average value of any phase function
\cite{Grabert}. In particular, for the relevant variable in
(\ref{relevant_variable}), the theory produces the following transport equation
for the average value $\left<F\right> = f$,
\begin{equation}
\label{dfdt}
\frac{\partial f}{\partial t} = -v(f)
                                             + \int_0^t K(t,\ s)\ ds.
\end{equation}
The first term on the right is known as the organized drift, $v$, while $K$ is
called the after-effect function. Here we will concentrate only on the
organized drift, defined as
\begin{equation}
\label{organised_drift}
v(f) =
  \mathrm{Tr}\left[\bar{\rho}\hat{\mathbf{L}} F\right] =
  \mathrm{Tr}\left[\bar{\rho}\sum_{i=1}^N\delta(q_i)\frac{p_i}
                                                         {m_i}\right],
\end{equation}
as a first crude approximation to the time rate of change for $f$. That is to
say, we are assuming that $K(t,\ s) \approx 0$. In (\ref{organised_drift}) we
have applied the Liouville operator $\hat{\mathbf{L}}$ to $F$ to calculate how
$v$ is related to the momenta $p_i$ and masses $m_i$ of the particles.
\begin{equation}
\hat{\mathbf{L}}F = \{H,\ F\} = \sum_{i=1}^N\delta(q_i)\frac{p_i}{m_i}
\end{equation}
If we follow Jaynes's maximum entropy route to determine the relevant
distribution $\bar{\rho}$ for $f$ equal to some value $N_R$, then we get
\begin{equation}
\label{doublewell_rhobar}
\bar{\rho}(z) = \frac {\delta(H(z)-E) e^{-\lambda_{N_R}F(z)}}
                      {\mathrm{Tr}\left[\delta(H(z)-E)
                                        e^{-\lambda_{N_R}F(z)}\right]},
\end{equation}
with the Lagrange multiplier $\lambda_{N_R}$ chosen to satisfy the constraint
on the average value of $F$. When we insert (\ref{doublewell_rhobar}) in
(\ref{organised_drift}) and integrate, the organised drift vanishes because
it is the integral of an odd function, due to the presence of the momenta
$p_i$.
\begin{equation}
\label{no_drift}
v(f) = 0.
\end{equation}

Surprisingly, in many cases this conclusion (\ref{no_drift}) is
incorrect. Suppose we choose an initial state with all the particles on the
left. If the average energy per  particle $E/N$ lies below the height of the
potential barrier, then the system can never  reach a state with all the
particles on the right, simply because there is not enough energy  to get them
all over the barrier. Note, though, that when we switch the signs of all the 
coordinates in our initial state we create a new inaccessible state with the
same total  energy as before. In other words, the system is not ergodic for
some values of the energy  $E$, and so it does not explore the complete
$H(z)=E$ surface in phase space.

Let $\rho^{ref}$ designate the final stationary distribution reached by the
system, to distinguish it from the microcanonical $\rho^{eq}$ implied by the
maximum entropy approach. Calculating the relevant distribution from the
relative entropy functional (\ref{Def-Relative_entropy}), the expression for
the organised drift becomes
\begin{equation}
v(N_R) =
  \frac{\mathrm{Tr}\left[{\rho^{ref}e^{-\lambda_{N_R}F}}
                   \hat{\mathbf{L}}F\right].}
  {\mathrm{Tr}\left[\rho^{ref}e^{-\lambda_{N_R}F}\right]}
\end{equation}
Even though $\rho^{ref}$ is unknown, we can sample it by means of a molecular
dynamics simulation. After the simulation run, the system has traversed a set
of points $\{z_j\}$, so we estimate $v(N_R)$ with
\begin{equation}
\label{average_drift}
v(N_R) \approx \frac{\sum_j e^{\lambda_{N_R}F(z_j)}\hat{\mathbf{L}}F(z_j)}
                    {\sum_j e^{\lambda_{N_R}F(z_j)}}.
\end{equation}

\begin{figure}
  \centering
    \includegraphics[width=\textwidth]{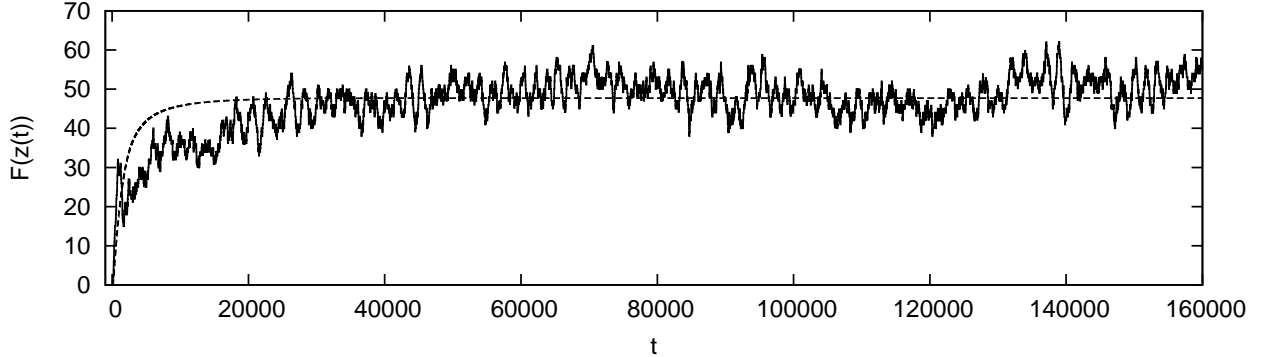}
    \caption{\label{simulation_run} Number of particles with a positive
      coordinate value versus time for a system of $N=100$ Lennard-Jones
      particles initially to the left of the potential barrier in figure
      \ref{doublewell}. A classic  Runge-Kutta fourth order algorithm with time
      step equal to $10^{-4}$ was used to integrate the equations of motion.
      The total energy remained constant to four significant figures. Numerical
      integration with the organised drift from figure \ref{simulated_drift}
      allowed us to estimate the average value of $F$ as a function of time
      (dashed line).}
\end{figure}

Figure \ref{simulation_run} represents the number of particles on the right of
the double-well potential in figure \ref{doublewell} as a function of time. We
started the simulation with all the particles on the left and an average energy
per particle $E/N$ below the height of the potential barrier. The average
kinetic temperature calculated over the whole run ($1.6\cdot 10^9$ time steps)
equals $T_{kin}/k_B = 1.2\pm0.1$ (mean $\pm$ standard deviation). When the
particles on either side of the barrier were considered separately, the average
kinetic temperature remained the same, but the standard deviation doubled on
the right of the well $T_{kin,R}/k_B = 1.2\pm0.2$.

\begin{figure}
  \centering
    \includegraphics[height=0.5\textwidth]{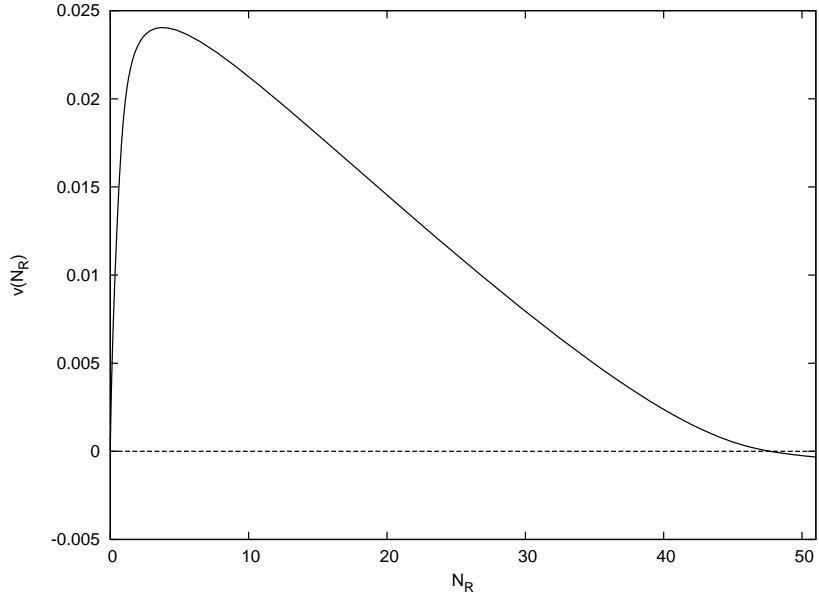}
    \caption{\label{simulated_drift} Organised drift versus average number of
       particles to the right of the potential barrier, $N_R$, calculated with
       equation (\ref{average_drift}).}
\end{figure}

Figure \ref{simulated_drift} shows the organised drift calculated with
(\ref{average_drift}) as a function of the average number of particles $N_R$ to
the right of the potential barrier. Numerical integration of equation
(\ref{dfdt}) by setting $K(t,s)=0$ generated the dashed line shown in figure
\ref{simulation_run}, which agrees qualitatively with the general trend of the
simulation. The deviations observed are not very surprising, considering that
we have completely neglected the after-effect function.

In summary, even though we have neglected the memory effects in the equation for
the organised drift (\ref{organised_drift}) calculated with the relative entropy
(as opposed to the Gibbs-Jaynes entropy method), we have achieved a very good
description of the transport over the energy barrier. The results are especially
interesting because the dynamical evolution took place under non-ergodic
conditions.

The method presented here could also be applied in principle to the numerical
calculation of the integral of $K(t,\ s)$ in (\ref{dfdt}), but this would involve
a much greater computational cost, so we have deferred these calcuations to
future research.

\section{Conclusions}
\label{conclusions}

In the context of the maximum entropy formalism, relative entropy
(\ref{Def-Relative_entropy}) has pleasant features. Its maximum value,
$\Delta S = 0$, obviously corresponds to equilibrium, and it enables us to
calculate the relevant distribution with less effort. Furthermore, the relevant
distribution turns into the equilibrium ensemble when the Lagrange multipliers
$\lambda_i$ vanish. With expressions like (\ref{rel_rho_2}) or
(\ref{relative_solution}) this fact lies in plain sight.

Given the equilibrium ensemble $\rho^{eq}$ and a set of constraints on
average values, $\left<F_i\right> = f_i$, the relevant distribution can
be calculated immediately by following these rules: first, write
\begin{equation}
\bar{\rho} = \frac{\rho^{eq}}{Z}e^{-\sum_i \lambda_i F_i}.
\end{equation}
Then ensure that $\bar{\rho}$ is normalised by writing
\begin{equation}
\mathrm{Tr}\left[\bar{\rho}\right] = 1
\end{equation}
and solving for the partition function $Z$. Finally, the $\lambda_i$ Lagrange
multipliers can be calculated, at least in principle, by inserting $\bar{\rho}$
into the constraints on the average values, $\left<F_i\right> = f_i$, and
solving for the $\lambda_i$.

The constraints for isolated systems must be considered carefully, because
knowledge of the equilibrium ensemble reveals the probability distribution
$P(i)$ over the whole set of values of the dynamical invariants through
\begin{equation}
\mathrm{Tr}\left[\rho^{eq}\delta(I - i)\right] = P(i).
\end{equation}
This information must be taken into account, so in this case we write
\begin{equation}
\bar{\rho} = \frac{\rho^{eq}}{\mathcal{Z}(i)}e^{-\sum_j \lambda_j F_j}.
\end{equation}
Then we ensure that $\bar{\rho}$ is normalised by writing
\begin{equation}
\mathrm{Tr}\left[\bar{\rho}\delta(I - i)\right] = P(i),
\end{equation}
and solving for $\mathcal{Z}(i)$. The relevant ensemble can then be used in
conjunction with the other constraints to find the value of the unknown
Lagrange multipliers.

The relative entropy method relies on our knowledge of the equilibrium ensemble,
and it provides no clues regarding how to calculate this probability
distribution, unlike Jaynes's method. However, this might also be interpreted as
a virtue. When the system has concealed dynamical invariants which have not been
taken into account, maximising the Gibbs-Jaynes entropy will not generally
reproduce the measured average values faithfully. This would signal the
existence of missing information. By contrast, if the equilibrium ensemble has
been determined by other means or if we are able to sample it effectively (with
molecular dynamics, for example), then we have simultaneously determined the
probability distribution for all the dynamical invariants. We may then simply
write the relevant distribution in terms of the equilibrium ensemble and, in
principle, use it to calculate nonequilibrium quantities like the coarse-grained
free energy or Green-Kubo transport coefficients \cite{Grabert}.

\section{acknowledgements}
We would like to express our gratitude to the anonymous reviewers of this
article for their insightful comments.


\end{document}